\documentclass{article}
\usepackage{spconf,amsmath,graphicx}
\usepackage{textcomp}
\usepackage{booktabs}
\usepackage{xcolor}
\usepackage{multirow}
\usepackage{subfigure}
\usepackage{booktabs}
\usepackage{fancyhdr}
\usepackage{marvosym}
\usepackage{setspace}
\title{A No-Reference Deep Learning Quality Assessment Method for Super-resolution Images  Based on Frequency Maps}
\name{ Zicheng Zhang(\Letter),Wei Sun,Xiongkuo Min,Wenhan Zhu,Tao Wang,Wei Lu,and Guangtao Zhai\address{Institute of Image Communication and Network Engineering, Shanghai Jiao Tong University, China \\
zzc1998@sjtu.edu.cn}}
\begin{document}
%\ninept
%
\maketitle
\let\thefootnote\relax\footnotetext{ This work was supported in part by NSFC 61901260 and NSFC 62101325.}

\begin{abstract}
To support the application scenarios where high-resolution (HR) images are urgently needed, various single image super-resolution (SISR) algorithms are developed. However, SISR is an ill-posed inverse problem, which may bring artifacts like texture shift, blur, etc. to the reconstructed images, thus it is necessary to evaluate the quality of super-resolution images (SRIs). Note that most existing image quality assessment (IQA) methods were developed for synthetically distorted images, which may not work for SRIs since their distortions are more diverse and complicated. Therefore, in this paper, we propose a no-reference deep-learning image quality assessment method based on frequency maps because the artifacts caused by SISR algorithms are quite sensitive to frequency information. Specifically, we first obtain the high-frequency map (HM) and low-frequency map (LM) of SRI by using Sobel operator and piecewise smooth image approximation. Then, a two-stream network is employed to extract the quality-aware features of both frequency maps. Finally, the features are regressed into a single quality value using fully connected layers. The experimental results show that our method outperforms all compared IQA models on the selected three super-resolution quality assessment (SRQA) databases.   
\end{abstract}
\begin{keywords}
super-resolution, no-reference image quality assessment, frequency maps, deep learning
\end{keywords}
\section{Introduction}
\label{sec:intro}
Single image super-resolution (SISR) algorithms are developed to reconstruct HR images from corresponding LR ones, which has been widely used in practical application scenarios like medical treatment, live stream media, etc. \cite{sr-application,sr2}. However, similar to other image enhancement operations, SISR algorithms inevitably introduce new artifacts such as ringing, texture shift, structural damage, and blur, etc. to the generated super-resolution images, which need to be evaluated quantificationally to promote the development of SISR algorithms.

\begin{figure}[tb]

\begin{minipage}[b]{.45\linewidth}
  \centering
  \centerline{\includegraphics[width=3.3cm]{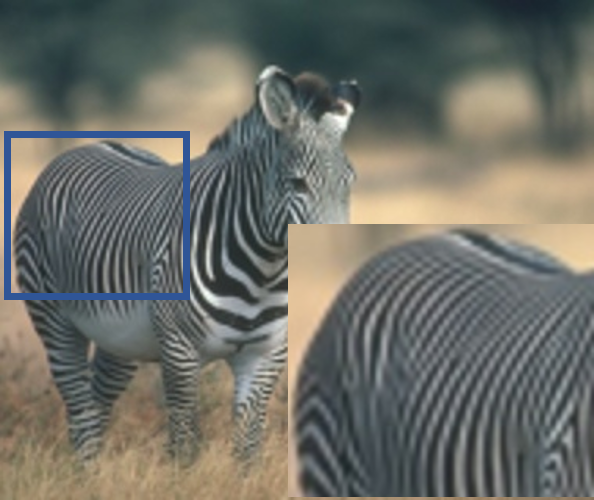}}
%  \vspace{2.0cm}
  \centerline{(a) \small High-frequency artifacts}\medskip
\end{minipage}
\begin{minipage}[b]{.58\linewidth}
  \centering
  \centerline{\includegraphics[width=3.3cm]{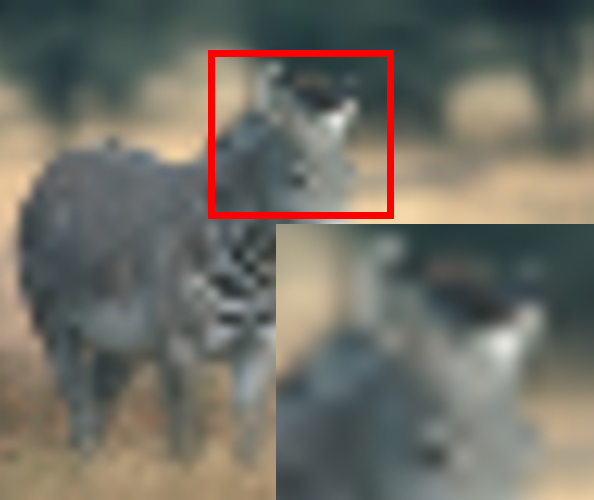}}
%  \vspace{1.5cm}
  \centerline{(b) \small Low-frequency artifacts}\medskip
\end{minipage}
\vspace{-1cm}
\caption{Typical examples of artifacts caused by SISR algorithms.}
\label{fig:dis}
\vspace{-0.6cm}
\end{figure}

\begin{figure*}[h]
    \centering
    \includegraphics[width=12cm]{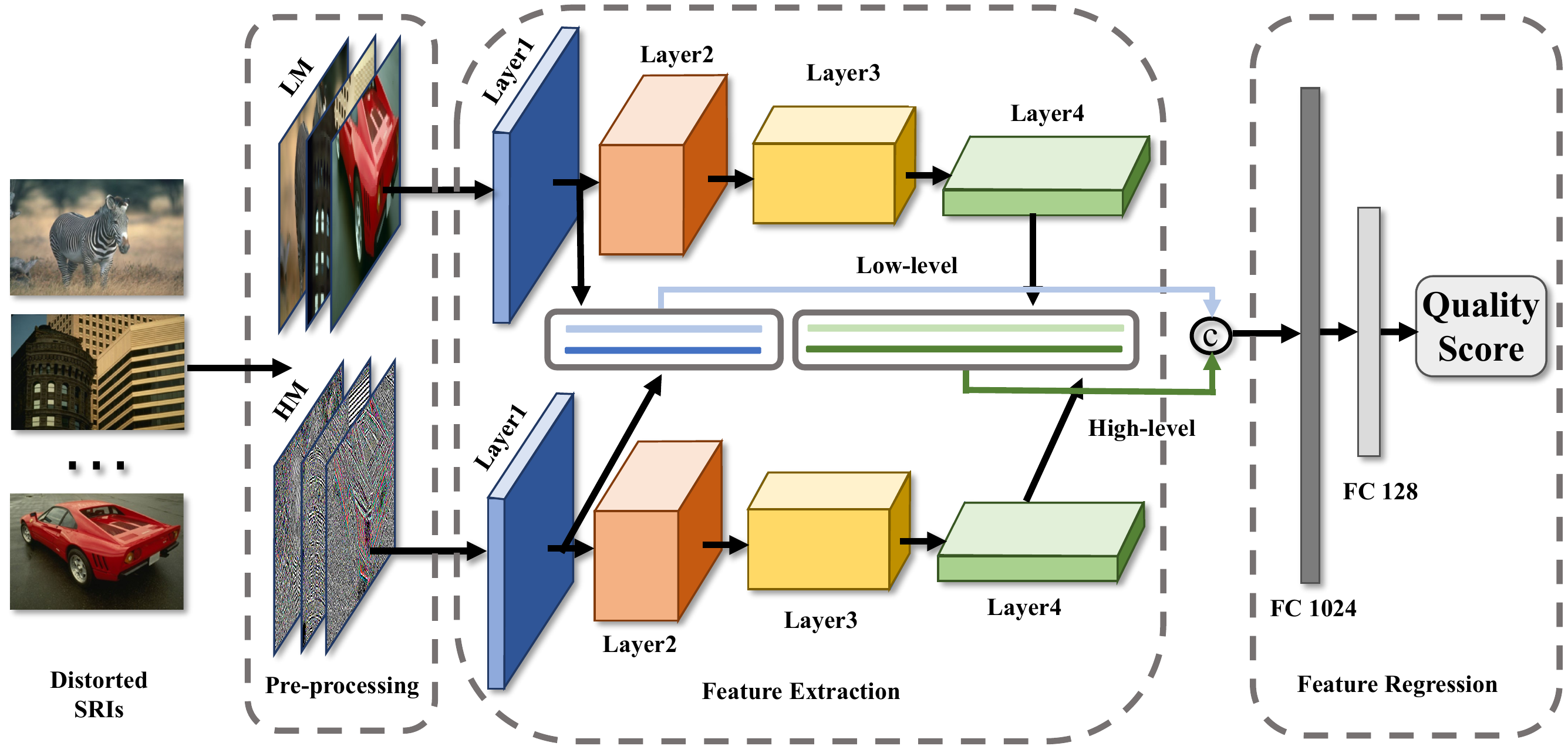}
    \caption{Framework of the proposed method with ResNet101 as the backbone. The two-stream networks do not share parameters.}
    \label{fig:framework}
    \vspace{-0.5cm}
\end{figure*}

In the current literature, objective image quality assessment (IQA) methods can be categorized into full-reference (FR) methods and no-reference (NR) methods according to the involvement of the reference images \cite{sun2}. Among FR-IQA methods, PSNR and SSIM \cite{ssim} are the two most widely used methods. However, they have been proven to have a low correlation with subjective judgment when the distortion is complex, especially for SRIs \cite{cviu}. Recently, to assess the quality of SRIs, Zhou $et$ $al.$ proposed to measure the changes of texture and structure distributions \cite{qads}. Using similar ideas, Wei $et$ $al.$ employed the structural fidelity and statistical naturalness to quantify the distortions \cite{sfsn}. However, the reference image is not always available, which encourages the development of NR-IQA models. Beron $et$ $al.$ proposed a statistic-based NR-IQA method for SRIs using optimal feature selection \cite{srij}. Supported by the success of convolution neural network (CNN), Fang $et$ $al.$ designed a CNN architecture to extract high-level intrinsic features for SRI quality assessment \cite{cnnsr}. Further, Wei $et$ $al.$ used the structure and texture information as the CNN input instead of directly learning from the distorted SRIs \cite{deepsrq}. 

In this paper, we propose a no-reference deep learning method for super-resolution image quality assessment (SRQA) based on frequency maps.  The distortions of SRIs can be categorized into high-frequency artifacts (like ringing and texture shift) and low-frequency damage (like structural damage and blur) as shown in Fig. \ref{fig:dis}. Therefore, we first generate the high-frequency map and low-frequency map by using Sobel operator and piecewise smooth image approximation \cite{lm}. Then we design a two-stream deep learning network to extract features from the frequency maps. ResNet \cite{resnet} is utilized as the feature extraction backbone and we also employ other CNN models such as VGG \cite{vgg} and MobileNet \cite{mobilenet} for comparison. To make full use of low-level and high-level features, we combine the output features of the first layer and last layer of ResNet as the quality-aware features. The obtained features are finally pooled through two-stage fully connected layers to get the quality score. To demonstrate the effectiveness of the proposed method, three publicly available SRQA databases (CVIU \cite{cviu}, QADS \cite{qads}, and SRIJ \cite{srij}) are selected for validation. Various state-of-the-art IQA models (both FR and NR) are chosen as the evaluation competitors. Further, the single-stream network ablation experiment and cross-database validation are conducted to test the effectiveness of the proposed method's structure. The experimental results show that the proposed method achieves the best performance among all the comparing IQA models on all three selected SRQA databases.

\vspace{-0.3cm}
\section{Proposed Method}
\label{sec:proposed_method}
The framework of the proposed method is shown in Fig. \ref{fig:framework}, which includes the pre-processing module, and the feature regression module.
\vspace{-0.1cm}
\subsection{Image Pre-processing}
In previous research, the distortions caused by SISR algorithms are divided into image structure and texture damage \cite{deepsrq}. In this paper, we further categorize the distortions into high-frequency artifacts (like ringing and texture shift) and low-frequency damage (like body structural damage and blur). 
Therefore, we employ the high-frequency maps and low-frequency maps of SRIs as the deep learning network inputs. Given image $I$, to generate the high-frequency map, we simply use the Sobel operators to sharpen the image details for enhancing the high-frequency artifacts components, which has been employed in many IQA tasks \cite{zhang2021ano,9675389}.
\begin{equation}
\boldsymbol{HM}=\sqrt{\left(I \otimes {S}_{x}\right)^{2}+\left(I \otimes {S}_{y}\right)^{2}}, 
\end{equation}
where $\boldsymbol{HM}$ indicates the high-frequency map, and ${S}_{x}$ and ${S}_{y}$ are the horizontal and vertical Sobel operators. To maintain the low-frequency damage and filter out the high-frequency artifacts' influence, we utilize the piecewise smooth image approximation (PSIA) proposed in \cite{psia} to generate the low-frequency maps:
\begin{align}
& \quad \quad \quad \quad \quad \quad \quad \boldsymbol{Minimize} \quad \boldsymbol{\tau} \\
&\boldsymbol{\tau}\! =\! \frac{1}{2} \! \int_{\Omega}\!(I-\boldsymbol{LM})^{2} \! d P\! + \!\beta \! \int_{\Omega \backslash K}\!\!\!|\nabla\! \boldsymbol{LM}|^{2} d P \!+ \!\alpha\! \int_{K}\! d \sigma
\end{align}
where $\boldsymbol{LM}$ denotes the low-frequency map, $\Omega$ is the image domain, $K$ denotes the edge set, $\int_{K} d \sigma$ represents the total edge length, $P$ indicates the pixel, and the coefficients $\alpha$ and $\beta$ are positive regularization constants.  In this way, the different types of distortions of SRIs can be separated into low-frequency artifacts and high-frequency artifacts, which improves the efficiency and effectiveness of the feature extraction module.

\begin{figure}[t]

\begin{minipage}[b]{.31\linewidth}
  \centering
  \centerline{\includegraphics[width=2.5cm]{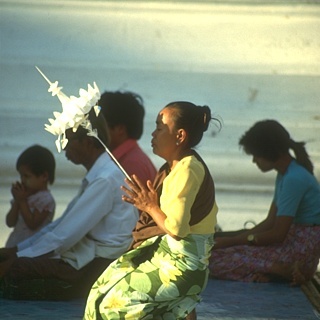}}
%  \vspace{2.0cm}
  \centerline{(a) \small Reference }\medskip
\end{minipage}
\begin{minipage}[b]{.36\linewidth}
  \centering
  \centerline{\includegraphics[width=2.5cm]{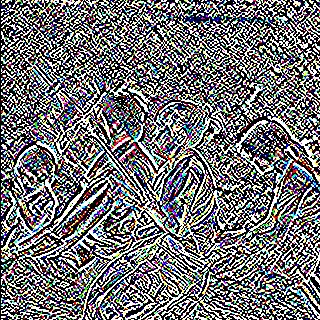}}
%  \vspace{1.5cm}
  \centerline{(b) \small Ref HM}\medskip
\end{minipage}
\hfill
\begin{minipage}[b]{0.31\linewidth}
  \centering
  \centerline{\includegraphics[width=2.5cm]{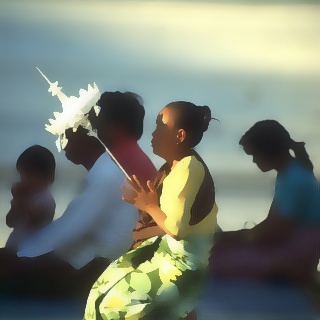}}
%  \vspace{1.5cm}
  \centerline{(c) \small Ref LM}\medskip
\end{minipage}
\begin{minipage}[b]{.31\linewidth}
  \centering
  \centerline{\includegraphics[width=2.5cm]{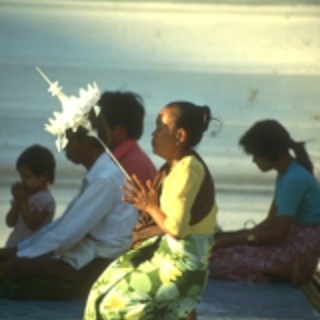}}
%  \vspace{2.0cm}
  \centerline{(d) \small SRI}\medskip
\end{minipage}
\begin{minipage}[b]{.36\linewidth}
  \centering
  \centerline{\includegraphics[width=2.5cm]{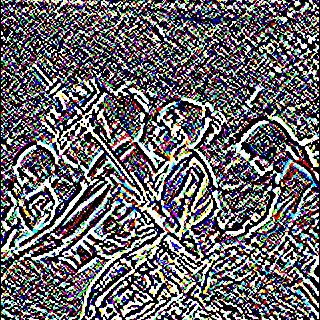}}
%  \vspace{1.5cm}
  \centerline{(e) \small SRI HM}\medskip
\end{minipage}
\hfill
\begin{minipage}[b]{0.31\linewidth}
  \centering
  \centerline{\includegraphics[width=2.5cm]{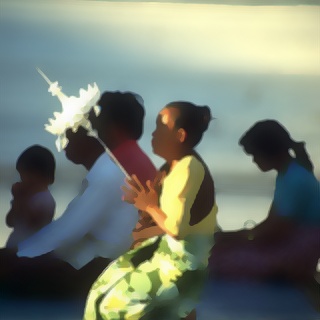}}
%  \vspace{1.5cm}
  \centerline{(f) \small SRI LM}\medskip
\end{minipage}
\vspace{-1cm}
\caption{Comparison of the reference image and its corresponding SRI from the CVIU \cite{cviu} database. HM and LM indicate the high-frequency map and low-frequency map respectively.}
\label{fig:res}
\vspace{-0.8cm}
\end{figure}

\subsection{Feature Extraction}
With the obtained high-frequency maps and low-frequency maps of distorted SRIs, we propose to use ResNet101 \cite{resnet} as the backbone, which has been widely used in many IQA studies \cite{sun3}. Then, we design a two-stream deep learning framework and the two-stream networks are independent, which indicates that they do not share parameters.  Considering that the features extracted from different layers of ResNet usually represent different visual information, we incorporate the feature maps captured from the first layer and last layer of ResNet101 as the quality-aware combined feature maps ($F$). Finally, the spatial global average pooling ($\rm GP_{mean}$) and spatial global standard deviation pooling ($\rm GP_{std}$) are employed to get the quality-aware feature vector from $F$:
\begin{equation}
    \begin{aligned}
    F_{mean} &= {\rm GP_{mean}}(F),\\
    F_{std} &= {\rm GP_{std}}(F),\\
    F_{quality} &= {\rm cat}(F_{mean},F_{std}),\\
    \end{aligned}
\end{equation}
where $F_{mean}$ and $F_{std}$ represent the global average and standard deviation results of $F$, $\rm cat(\cdot)$ indicates the concatenation operation, and $F_{quality}$ denotes the final quality-aware feature vector.

\vspace{-0.1cm}
\subsection{Feature Regression}
After the feature extraction module, two-stage fully connected layers consisting of 1024 and 128 neurons are utilized as the regression module. Then the predicted perceptual quality level $\boldsymbol{Q}_{predict}$ for SRIs can be computed as:
\begin{equation}
    \boldsymbol{Q}_{predict} = {\rm FC}(F_{quality}),
\end{equation}
where $\rm FC$ denotes the fully connected pooling operation. The mean squared error (MSE) is utilized as the loss function, which can be denoted as:
\begin{equation}
    Loss = || \boldsymbol{Q}_{predict}- \boldsymbol{Q}_{label}||^2_{2}
\end{equation}
where $\boldsymbol{Q}_{label}$ represent the mean opinion scores (MOSs) collected from subjective experiments.

\begin{table*}[t]
\renewcommand\tabcolsep{3pt}
\renewcommand\arraystretch{1.23}
\caption{Experimental results on the SRQA databases. The SRI*, HM*, and LM* represent the single-stream network of original SRIs, high-frequency maps, and low-frequency maps with ResNet101 as the feature extraction backbone. The Proposed(VGG), Proposed(Mob), and Proposed(Res) indicate the proposed two-stream network with VGG16, MobileNet, and ResNet101 as the feature extraction backbone.}
\vspace{-0cm}
\centering
\begin{tabular}{c|c|cccc|cccc|cccc}
\toprule
\multicolumn{1}{c|}{\multirow{2}{*}{Type}} & \multicolumn{1}{c}{\multirow{2}{*}{Mehtods}} & \multicolumn{4}{|c}{CVIU}                                              & \multicolumn{4}{|c}{QADS}                                              & \multicolumn{4}{|c}{SRIJ}                                              \\ \cline{3-14}
\multicolumn{1}{c|}{}  
&         & SRCC    & PLCC   & KRCC   & RMSE   & SRCC    & PLCC   & KRCC   & RMSE  & SRCC    & PLCC   & KRCC   & RMSE \\ \hline \multirow{7}{*}{FR} 
& PSNR    & 0.5696  & 0.5825 & 0.3969 & 2.7383 & 0.4052  & 0.4188 & 0.2797 & 2.2512 & 0.3792 & 0.3766 & 0.2562 & 2.6494 \\
& SSIM    & 0.6761  & 0.6903 & 0.5752 & 1.7980 & 0.5915  & 0.5915 & 0.4065 & 2.2325 & 0.4846 & 0.4553 & 0.3256 & 2.1369\\
& MS-SSIM & 0.7945  & 0.7443 & 0.5900 & 1.0272 & 0.6837  & 0.6837 & 0.4991 & 1.1896 & 0.5351 & 0.4427 & 0.3589 & 1.4421 \\
& FSIM    & 0.7485  & 0.7404 & 0.5454 & 1.2321 & 0.7026  & 0.7026 & 0.5153 & 1.1988 & 0.5355 & 0.4828 & 0.3600 & 1.1058 \\
& MAD     & 0.8658  & 0.8581 & 0.6697 & 1.0131 & 0.7380  & 0.7380 & 0.5427 & 1.1634 & 0.5144 & 0.5173 & 0.3530 & 1.9358\\
& GMSD    & 0.8463  & 0.8338 & 0.6490 & 1.1060 & 0.7391  & 0.7391 & 0.5423 & 1.1554 & 0.4790 & 0.4731 & 0.3205 & 2.1125\\
& STD     & 0.8661  & 0.8842 & 0.6797 & 1.1383 & 0.9120  & 0.9225 & 0.8861 & 0.8382 & 0.7780 & 0.7510 & 0.5561 & 1.0383 \\
& SFSN    & 0.8716  & 0.8981 & 0.6771 & 0.9980 & 0.8319  & 0.8319 & 0.6396 & 1.1459 & 0.7586 & 0.7347 & 0.5572 & 0.9126 \\ 
 \hline

%         & SRCC    & PLCC   & KRCC   & RMSE   & SRCC    & PLCC   & KRCC   & RMSE  & SRCC    & PLCC   & KRCC   & RMSE \\ \hline 
\multirow{11}{*}{NR} 
& BRISQUE & 0.7255  & 0.5825 & 0.6521 & 1.7383 & 0.8052  & 0.8143 & 0.7775 & 1.2383 & 0.7164 & 0.7270 & 0.6521 & 1.4353 \\
& NIQE    & 0.6231  & 0.5825 & 0.4375 & 1.5582 & 0.5837  & 0.5945 & 0.3975 & 2.1375 & 0.4913 & 0.5560 & 0.4375 & 2.7565\\
& OFS     & 0.8472  & 0.8318 & 0.6510 & 1.2126 & 0.7571  & 0.5825 & 0.5969 & 1.3120 & 0.8630 & 0.8580 & 0.6786 & \textbf{0.7317} \\ 
& CNN-SR  & 0.8394  & 0.9156 & -      & 1.2527 & 0.8541  & 0.8709 & -      & 1.5080 & -      & -      & -      & -      \\
& DeepSRQ & 0.9206  & 0.9273 & -      & 0.9042 & 0.9528  & 0.9557 & -      & 0.7670 & -      & -      & -      & -      \\
& HM* & 0.9252  & 0.9330 & 0.7500 & 0.8048 & 0.9205  & 0.9389 & 0.7833 & \textbf{0.7402} & 0.8796 & 0.8889 & 0.6134 & 0.8598 \\
& LM* & 0.9364  & 0.9494 & 0.7969 & 0.8008 & 0.9164  & 0.9199 & 0.7166 & 0.7796 & 0.8664 & 0.8669 & 0.6662 & 0.8189 \\
& SRI*& 0.9385 & 0.9273 & 0.8000 & 0.7900 & 0.9588  & \textbf{0.9514} & \textbf{0.8333} & 0.7552 & 0.8772 & 0.8823 & 0.6874 & 0.8098 \\
& Proposed(VGG)    & 0.9454  & 0.9418 & 0.7896 & 0.8068 & 0.9452  & 0.9302 & 0.7733 & 0.8071 & 0.8663 & 0.8713 & 0.6480 & 0.9497 \\
& Proposed(Mob)& 0.9147& 0.8933 & 0.7833 & 0.8224 & 0.9252  & 0.9149 & 0.7335 & 0.8421 & 0.8393 & 0.8451 & 0.6431 & 0.9116 \\
& Proposed(Res)  & \textbf{0.9478}  & \textbf{0.9475} & \textbf{0.8091} & \textbf{0.7752} & \textbf{0.9613}  & 0.9504 & 0.8294 & 0.7414 & \textbf{0.8887} & \textbf{0.8909} & \textbf{0.6993} & 0.7547 \\

\bottomrule
\end{tabular}
\label{tab:performance}
\vspace{-0.6cm}
\end{table*}

\vspace{-0.1cm}
\section{Experiment}
\subsection{Experiment Databases and Criteria}
To test the effectiveness of the proposed method, we select 3 SRQA databases for validation, which include CVIU \cite{cviu}, QADS \cite{qads}, and SRIJ \cite{srij}. Specifically, CVIU includes 1,620 color SRIs generated from 30 LR images, QADS includes 980 color SRIs generated from 20 LR images, and SRIJ includes 608 gray SRIs generated from 32 gray LR images. For validation on SRIJ, we simply pass the gray information to RGB channels to fit the deep learning network.

Four mainstream consistency evaluation criteria are utilized to compare the correlation between the predicted scores and MOSs, which include Spearman Rank Correlation Coefficient (SRCC), Kendall’s Rank Correlation Coefficient (KRCC), Pearson Linear Correlation Coefficient (PLCC), Root Mean Squared Error (RMSE). Specifically, when calculating RMSE, we uniformly scale the MOSs of different databases and predicted scores of different IQA models to 0-10 points for comparison.

\begin{table}
\caption{Experimental results of cross-database validation.}
%\vspace{0.2cm}
\renewcommand\tabcolsep{7.5pt}
\renewcommand\arraystretch{1}
    \centering
    \begin{tabular}{c|cc|cc}
    \toprule
           Train$\xrightarrow{}$Test & \multicolumn{2}{c|}{CVIU$\xrightarrow{}$QADS} &\multicolumn{2}{|c}{QADS$\xrightarrow{}$CVIU} \\ \hline
            Method &SRCC &PLCC &SRCC &PLCC \\ \hline
           OFS  & 0.7520    & 0.7470 & 0.7622 & 0.7753    \\ 
           DeepSRQ & 0.7225    & 0.7486  &- &-\\ 
           Proposed &\textbf{0.7863} &\textbf{0.7725} &\textbf{0.7845} &\textbf{0.7757} \\
    \bottomrule
    \end{tabular}
    \vspace{-0.6cm}
    \label{tab:cross}
\end{table}
\vspace{-0.1cm}
\subsection{Experiment Setup}
Considering different sizes of SRIs from different SRQA databases, we randomly crop the SRIs into patches with sizes of 320x320, 380x380, and 448x448 for CVIU, QADS, and SRIJ databases during each epoch.  We use the Adam optimizer with initial learning rate set as 1e-4 and set the batch size as 16. Because the proposed method requires a training process, we randomly split each database into the training set with about 80\% SRIs and the testing set with about 20\% SRIs. There is no overlap of the same image content between the training and testing sets. We repeat the split process 10 times and record the average performance as final experimental results.

Except for ResNet101, other CNN models such as VGG16 and MobileNet are employed as the backbone to test the influence of different feature extraction models. Furthermore, we also test the performance of single-stream network as the ablation experiment.
\vspace{-0.1cm}
\subsection{Evaluation Competitors}
To demonstrate the effectiveness of the proposed method, some classic IQA methods along with some deep learning methods are selected as competitors, which can be categorized into two types respectively:
\\ $\bullet$   FR-IQA models: PSNR, SSIM \cite{ssim}, FSIM \cite{fsim}, MAD \cite{mad}, GMSD \cite{gmsd}, STD \cite{qads}, and SFSN \cite{sfsn}, among which STD and SFSN are specially designed for dealing SRQA problems while others are IQA models developed for general IQA problems.
\\ $\bullet$  NR-IQA models: BRISQUE \cite{brisque}, NIQE \cite{niqe}, OFS \cite{srij}, CNN-SR \cite{cnnsr}, DeepSRQ \cite{deepsrq}, among which OFS, CNN-SR, and DeepSRQ are specially developed for SRQA tasks while others are carried out for general IQA problems. What's more, CNN-SR and DeepSRQ are deep-learning based methods.

\vspace{-0.1cm}
\vspace{-0.1cm}
\subsection{Experiment Performance}

The experiment performance is clearly shown in Table \ref{tab:performance}, and the best performance result is marked in bold for each column. We can see that the proposed method using ResNet101 outperforms all the comparing IQA models in SRCC of all three databases, which indicates that our method has a good ability to accurately predict the quality ranks of SRIs. Furthermore, by comparing the single-stream network results and the proposed two-stream network results, we can find that 
Proposed(Res) achieves better performance than SRI*, HM*, and LM*, which confirms the contributions of two frequency maps to the final results and validates the structure of the proposed method.

To further test the generalization ability of the proposed model, we also give the results of cross-database validation in this section. Additionally, because SRIJ database contains only gray images, CVIU and QADS databases are chosen. To deal with the problem of size mismatch between CVIU and QADS databases, we retrain the Proposed(Res) model on QADS database by cropping the SRIs into 320x320 patches. 
The SRCC and PLCC results of cross-database validation are listed in Table \ref{tab:cross}. The best result is marked in bold in each row. It can be observed that the proposed method obtains better generalization ability compared with other NR-IQA models.

\vspace{-0.5cm}
\section{Conclusion}
\vspace{-0.3cm}

In this study, we build an effective deep learning NR metric based on frequency maps for super-resolution image quality assessment problems. We generate the high-frequency map and low-frequency map by using Sobel operator and piecewise smooth image approximation. A two-stream deep learning network is designed to extract features of both frequency maps. Finally, the features are regressed into quality scores by using two-stage fully connected layers. The experimental results show that our method gains better performance comparing with the other existing state-of-art IQA models and obtains good stability among all the three selected SRQA databases. The cross-database validation further confirms the effectiveness and generalization ability of the proposed method.

% References should be produced using the bibtex program from suitable
% BiBTeX files (here: strings, refs, manuals). The IEEEbib.bst bibliography
% style file from IEEE produces unsorted bibliography list.
% -------------------------------------------------------------------------
\bibliographystyle{IEEEbib}
\bibliography{strings,refs}

\end{document}